\documentclass{JBCS-mod}
\usepackage[T1]{fontenc}
\usepackage{algorithm}
\usepackage{subfigure}

\usepackage{ifpdf}
\ifpdf
\usepackage[pdftex]{graphicx}
\else
\usepackage[dvips]{graphicx}
\fi

\usepackage[english,brazil]{babel}

\usepackage[latin1]{inputenc}

\usepackage{url}

\begin{document}

\setcounter{page}{01}

\cabnames{E. Valadão, D. Guedes, R. Duarte}
\cabtit{Caracter. de tempos de ida-e-volta na Internet}

\title{Caracterização de tempos de ida-e-volta na Internet}


\namea{Everthon Valadão$^1$, Dorgival Guedes$^2$, Ricardo Duarte$^3$}


\addressa{
\\
  $^1$Departamento de Informática\\
  Instituto Federal de Educação, Ciência e Tecnologia de Minas Gerais\\
  Campus Formiga\\
  Rua Padre Alberico, s/n --- Bairro São Luiz\\
  CEP 35570-000, Formiga, MG\\
  Telefone: (037) 3321-4094 Fax (037) 3322-2330\\
  everthon.valadao@ifmg.edu.br\\
\\
  $^2$Departamento de Ciência da Computação\\
  $^3$Departamento de Engenharia Eletrônica\\
  Universidade Federal de Minas Gerais\\
  Av. Antônio Carlos 6627 --- Pampulha\\
  CEP 31270-010,  Belo Horizonte, MG\\ 
  Telefone: (031) 3409-5860 Fax: (031) 3409-5858\\ 
  dorgival@dcc.ufmg.br\\
  ricardoduarte@ufmg.br 
}

\nameb{}
\addressb{}

\namec{}
\addressc{}

\named{}
\addressd{}


\maketitle

\begin{abstract}
  Round-trip times (RTTs) are an important metric for the operation of many
applications in the Internet.
For instance, they are taken into account when choosing
servers or peers in streaming systems, and they impact the operation of fault
detectors and congestion control algorithms.
Therefore, detailed knowledge about RTTs is important for
application and protocol developers.
In this work we present results on measuring RTTs between 81
PlanetLab nodes every ten seconds, for ten days.
The resulting dataset has over 550 million measurements.
Our analysis gives us a profile of delays in the network
and identifies a Gamma distribution as the model that best fits our data.
The average times observed are below 500~ms in more than 99\% of the pairs,
but there is significant variation, not only when we compare different
pairs of hosts during the experiment, but also considering any given pair
of hosts over time.
By using a
clustering technique, we observe that links can be divided in five 
distinct groups
based on the distribution of RTTs over time and the losses observed,
ranging from groups of near, well-connected pairs, to groups of distant
hosts, with lower quality links between them.

\end{abstract}
\pagebreak

\begin{resumo}
  Tempos de ida-e-volta (RTTs) são uma métrica importante
para diversas aplicações na Internet.
Por exemplo, eles afetam
a escolha de servidores ou pares em sistemas de troca de
arquivos e de \textit{streaming} e a operação de mecanismos de
detecção de falhas e de controle de congestionamento.
Por isso, entender a variação de RTTs 
é essencial para o projeto de protocolos e aplicações.
Neste trabalho apresentamos os resultados da medição de RTTs
entre 81 nós da rede PlanetLab a cada dez segundos, por dez dias.
O registro resultante possui cerca de 550 milhões de medições.
Nossa análise nos permite traçar um perfil detalhado do atraso
na rede e identificar uma distribuição Gamma como a que melhor
aproxima os dados coletados.
Os tempos médios observados são inferiores a 500~ms em mais de 99\% dos
casos, mas a variação é significativa, não só para diferentes
pares de máquinas, mas também
para um mesmo par ao longo do tempo.
Aplicando uma técnica de
clusterização, verificamos que os enlaces podem ser
classificados em cinco grupos com base na distribuição dos RTTs,
sua variação e perdas, variando de grupos de nós próximos 
conectados a grupos de nós distantes com enlaces de menor qualidade.

\end{resumo}
\palavraschave{Tempos de ida-e-volta, RTT, atrasos, Internet, PlanetLab}

\setcounter{footnote}{0}

\section{Introdução}
\label{sec:introducao}

Na Internet atual, muitas aplicações dependem de informações sobre os
tempos de ida-e-volta pela rede (\emph{Round-Trip Times}, ou RTT) para
certas decisões. Além de ser essencial para se prever o comportamento de
protocolos como o TCP, que dependem de estimativas para esse
valor~\cite{karn1988,aikat2003},
esses tempos são importantes para aplicações que operam
segundo o princípio de requisição resposta e que podem escolher entre
diversos servidores na rede~\cite{ratnasamy2002},
aplicações de \emph{multicast} de
áudio/vídeo dependem dessa informação para montar suas árvores de
distribuição~\cite{kim2006}
e aplicações par-a-par, que podem se beneficiar ao
escolher como vizinhos nós que respondem mais rápido a consultas~\cite{choffnes2010network}.
Além disso, essa
informação também é usada quando se objetiva reduzir as latências de 
transmissão, dimensionar \emph{buffers} na rede e avaliar a
banda disponível entre dois
\emph{hosts}~\cite{wille2008,prasad2003bandwidth}.

Um estudo sobre a distribuição de atrasos (unidirecionais ou
de ida-e-volta) é importante para servir como base para auxiliar projetistas
de protocolos e aplicações em seu trabalho, já que fornece dados de
referência para a tomada de decisões. Além disso, os dados coletados
podem servir como entrada realista para simulações de sistemas
distribuídos e como base para o desenvolvimento de modelos sintéticos
a serem usados em simuladores.

Segundo Redígolo e outros, dentre os mecanismos de obtenção de
conhecimento, o método científico mostra-se, não apenas como o de êxito
mais provável, mas como o mais
convincente~\cite{redigolo-minicursoplanetlab}. O empirismo, cerne do método
científico, consiste na investigação controlada de um fenômeno, organizada
na forma de experimentos. Com efeito, os resultados positivos de um
experimento prestam-se ao intuito de comprovar ou ao menos, reforçar uma
hipótese formulada e investigada na condução deste. 

Através da utilização de um \textit{trace} na análise de protocolos de
rede, possibilita-se a investigação controlada e a repetibilidade dos
experimentos, de maneira que comparações de desempenho possam ser
realizadas num cenário real. Além disso, uma carga real possibilitaria uma
análise de desempenho considerando também efeitos inesperados, resultantes
da interação entre sistemas complexos, produzindo resultados com alto valor
científico e válidos como cenários reais.

Tendo esses objetivos em mente, neste trabalho apresentamos um estudo
sobre tempos de ida-e-volta e atrasos unidirecionais na
Internet. Medidas de RTT foram coletadas entre aproximadamente 100
pontos da rede, a cada dez segundos, durante dez dias. Os dados coletados
foram analisados utilizando ferramentas estatísticas e de mineração de
dados, buscando fornecer uma caracterização detalhada dos dados
coletados. 

Analisamos, entre outros aspectos, a distribuição de atrasos e perdas
observadas, o nível de assimetria observado durante as medições e a
existência de agrupamentos de canais de comunicação com características
distintas. Além da coleta, análise e disponibilização do \textit{trace} de
um cenário real, desenvolvemos também um modelo sintético para a
distribuição de RTTs.

Para uma medida precisa de tempos de ida-e-volta e de atrasos
unidirecionais é necessário acesso aos dois extremos do canal de
comunicação que se deseja medir, o que descarta o uso de aplicações
usuais que utilizam o protocolo ICMP,
como o comando {\tt ping}~\cite{luckie2001}. Sendo
assim, utilizamos nós da rede PlanetLab em nossas medições, tendo
o cuidado de selecionar máquinas com uma ampla cobertura geográfica,
localizadas tanto em redes acadêmicas/de pesquisa quanto em redes
comercias. 

Nossos resultados mostram que os RTTs variam significativamente entre pares
de máquinas na Internet, como seria de se esperar, mas também podem
variar significativamente entre duas máquinas definidas, mesmo em
escalas de tempo reduzidas. Apesar da variabilidade, nossa análise
indica que há alguns grupos de comportamento similar definidos
com base nos tipos de canais que as separam.
Perdas de pacotes, em geral, são baixas e tendem a se concentrar 
nos canais com maiores RTTs.

O restante deste trabalho é organizado da seguinte forma: a seção
a seguir discute conceitos e trabalhos relacionados, para então
discutirmos o processo de medição e coleta de dados na
seção~\ref{sec:coleta}. Os resultados são apresentados na
seção~\ref{sec:resultados}, com subseções para os principais elementos
da nossa análise. Finalmente, concluímos com algumas observações 
e uma discussão de possíveis trabalhos futuros.


\section{Trabalhos relacionados}
\label{sec:relacionados}

A avaliação do desempenho de um sistema (computacional) é significativa
somente no contexto de uma carga de trabalho, onde a utilidade de um modelo
de carga depende do quão bem o modelo sintético representa o efeito
observado em cargas reais e, se ele tem poder preditivo. Cargas são válidas
na medida em que permitem avaliar corretamente o comportamento do sistema
no que diz respeito às características de interesse.
Segundo Smith~ \cite{smith2007}, na medida em que o desempenho
é sensível a mudanças na
carga de trabalho, para outros cenários de carga as estimativas de
desempenho serão pouco confiáveis ou então o modelo ou metodologia de
estimativa serão considerados suspeitos. Ou seja, os resultados de
desempenho obtidos para uma determinada carga de trabalho são válidos para
o contexto daquela carga.

Choi e Yoo propuseram uma técnica para derivar atrasos
unidirecionais avaliando os tempos entre pacotes de dados e suas confirmações (ACKs)
em conexões TCP. Essa técnica foi analisada pelos autores através de
simulações, porém, não chegaram a aplicá-la na análise de
assimetria ou em um estudo abrangente na Internet~\cite{choi2005}. 

	Por exigir a medição nos dois extremos da comunicação e demandar
mecanismos de sincronização de relógios nesses extremos,
estudos sistemáticos sobre RTT (\textit{round-trip time}) na Internet
no passado foram limitados pela dificuldade de obter acesso a um número
significativo de máquinas instaladas em pontos diversos da rede. Por
exemplo, Bolot~\cite{bolot93} realizou medições apenas entre
um par de máquinas na rede, às quais o autor tinha acesso. Naquele
trabalho, o autor realizou medições por intervalos curtos (até 800 medições
consecutivas), variando o intervalo entre medições de 8~ms a 500~ms.

Trabalhos de medição de características da rede
só se tornaram realmente viáveis
recentemente, com o advento do
PlanetLab~\footnote{\url{http://www.planet-lab.org/}}.
O \emph{All-Pairs Ping}\footnote{\url{http://pdos.csail.mit.edu/~strib/pl_app/}} (APP)
foi um projeto que mediu RTTs no PlanetLab a cada 15
minutos por 2 anos~\cite{planetlab-app,planetlab-appv2}. 
Nesse caso o comando \textit{ping} foi utilizado, o que não
permite o casamento preciso entre as informações nas duas extremidades
do canal de comunicação. 
Já o serviço CoMon
realiza a monitoração do RTT, além de diversas métricas de carga, mas
não armazena a relação entre os pares de nós utilizados e se limita a um
intervalo de 5 minutos~\cite{park06comon}.
Numa análise do progresso evolutivo do RTT 
médio do PlanetLab~\cite{tang07-empiricalStudyPlab}, 
Tang e outros observaram uma
considerável variação no RTT entre duas medições para os mesmos pares de máquinas,
ao realizar medições apenas a cada 15 minutos.

Ainda no PlanetLab, Pucha e outros utilizaram pacotes formados como segmentos
de confirmação do protocolo TCP (ACKs) enviados para portos altos de máquinas
distantes como forma de medir RTTs. Pela definição do protocolo TCP, as máquinas
contactadas enviavam segmentos de erro (RSTs) de volta para a máquina de origem,
o que lhes permitia medir os tempos de ida-e-volta~\cite{pucha2006}. 

Em uma análise mais recente da assimetria dos atrasos na Internet,
Pathak e outros
realizaram um estudo abrangente no qual mediram a severidade da assimetria no
atraso entre diversos pares de nós do PlanetLab~\cite{pathak2008measurement}.
Seus resultados estão de acordo com os de Tang e outros, sendo observado que a
assimetria dos atrasos prevalece na grande maioria dos nós, podendo ser
em parte atribuída à assimetria nas rotas e em parte a congestionamentos
transientes.

Um resumo comparativo dos principais trabalhos de análise dos tempos de ida-e-volta em cenários da Internet e no PlanetLab é apresentado na Tabela~\ref{tab:comparacao-trabalhos}.

\begin{table*}[htb!]
        \caption{Sumário comparativo das pesquisas sobre atrasos na Internet} \label{tab:comparacao-trabalhos}	
	\begin{center}
	\begin{scriptsize}		
	\begin{tabular}{|l|c|c|c|c|c|c|c|}
\hline
\em \textbf{Trabalho} & \em \textbf{Método} & \em \textbf{Espaçamento / Duração} & \em \textbf{Variação Atraso} & \em \textbf{Assimetria} & \em \textbf{Modelo Sintético} & \em \textbf{Escala} \\
\hline
Bolot~\cite{bolot93}          & \emph{NetDyn} & 8--500 ms / 800 medições & $\surd$ & - & -  & 2 nós  \\
Choi e Yoo~\cite{choi2005} 		& TCP ACKs & $\sim$1 ms / 100 s & $\surd$ & - & -  & 2 nós  \\
Pucha e outros~\cite{pucha2006} & TCP ACK/RST  & ? & $\surd$ & -  & -  &  180 nós  \\
Yoshikawa~\cite{planetlab-appv2}		& \textit{ping}  & 15 min.
/ 2 anos & $\surd$ & - & -  & 600 nós  \\
Park e Pai~\cite{park06comon} 		& \textit{ping}  & 5 min. / - &
$\surd$ & - & -  & -  \\
Tang e outros~\cite{tang07-empiricalStudyPlab} 		&  \textit{ping}  & 15 min. / 2 anos & $\surd$ & - & - &  600 nós  \\
Pathak e outros~\cite{pathak2008measurement} 	& \emph{OwPing}  & 20 min. / 10 dias & $\surd$ & $\surd$ & - &  94 nós  \\
\textbf{Este trabalho} & \emph{3-way ping}  & \textbf{10 seg. / 10 dias} & \textbf{$\surd$} & \textbf{$\surd$} & \textbf{$\surd$} & \textbf{81 nós}  \\
\hline
	\end{tabular}
	\end{scriptsize}
	\end{center}
\end{table*}

O uso do PlanetLab para experimentos sobre a Internet é abordado
por Pathak e outros~\cite{pathak2008measurement} e por Pucha e outros~\cite{pucha2006}.
Ambos observaram
uma diferença de conectividade entre nós da Internet acadêmica e da
Internet comercial. Levamos esse aspecto em consideração ao incluir nós
localizados em redes comerciais em nossas medições.
Neste trabalho, entretanto, realizamos uma
análise de agrupamentos baseada nos valores medidos, não em uma classficação
\emph{a priori} dos nós.

Para medir atrasos unidirecionais, utilizamos uma técnica
derivada de Pathak e outros~\cite{pathak2008measurement}
que se baseia no uso do protocolo
NTP~\footnote{\url{http://www.ietf.org/rfc/rfc1305.txt}
} 
para sincronização dos relógios das máquinas e avaliação dos erros máximos
envolvidos. Rocha e outros discutem as limitações dessa
técnica~\cite{rocha2004}, especialmente os problemas de atualização de relógio, que foram consideradas
ao se excluir 
nós problemáticos, conforme será discutido posteriormente.

Como pode ser observado na tabela~\ref{tab:comparacao-trabalhos},
talvez por restrições da capacidade de coleta e processamento dos dados,
a maior parte dos projetos desenvolvidos para medir atrasos na
Internet foram baseados em períodos longos entre amostragens, da ordem de minutos,
ou realizaram medições por intervalos curtos de tempo. Entretanto, alguns desses
trabalhos indicam que 
as amostras muito espaçadas perdem detalhes da variação dos atrasos em intervalos
de tempo menores~\cite{tang07-empiricalStudyPlab}, 
enquanto medições muito próximas (da ordem do valor do RTT) podem
sofrer de interferências umas das outras~\cite{bolot93}.
Neste trabalho, optamos por adotar uma resolução temporal intermediária
(uma medição a cada 10 segundos) para evitar os dois problemas
mencionados. Além disso, utilizamos uma infra-estrutura de coleta contínua
para permitir que cada nó medidor pudesse operar por um período de tempo
mais longo (10 dias).

\section{Metologia de coleta}
\label{sec:coleta}

O RTT (\textit{round-trip time}) entre dois nós A e B é definido como
sendo a soma do atraso direto de A para B e do atraso inverso de B para
A. Ele tem sido utilizado largamente como métrica em diversas aplicações
de rede que necessitam inferir atrasos e/ou localização dos nós.
O RTT é comumente utilizado para derivar os atrasos em cada sentido, 
onde normalmente assume-se seu valor como sendo RTT/2. 
Entretanto, a existência de assimetrias nesses atrasos individuais
prejudica a precisão e desempenho das aplicações que se utilizam desse
pressuposto~\cite{pathak2008measurement}. Portanto, se possível, a 
utilização dos atrasos individuais em cada sentido do enlace é desejável 
por trazer consigo maior exatidão.

\subsection{Medição dos tempos de ida-e-volta}
\label{sec:medicao}
	
Para a realização da coleta dos atrasos de rede utilizamos uma ferramenta
com um protocolo próprio, visto que o \textit{ping}
(\textit{ICMP echo request}) não nos proveria as informações necessárias
para inferir os atrasos individuais de cada sentido da comunicação. Optamos
por utilizar uma abordagem ``\textit{Three-Way Ping}'', inspirada no
\emph{three-way handshake} utilizado no estabelecimento de conexões TCP,
de maneira a ter acesso ao RTT nos dois sentidos em momentos próximos.
	
\begin{figure}[h]
 \centering
 \includegraphics[width=0.7\columnwidth]{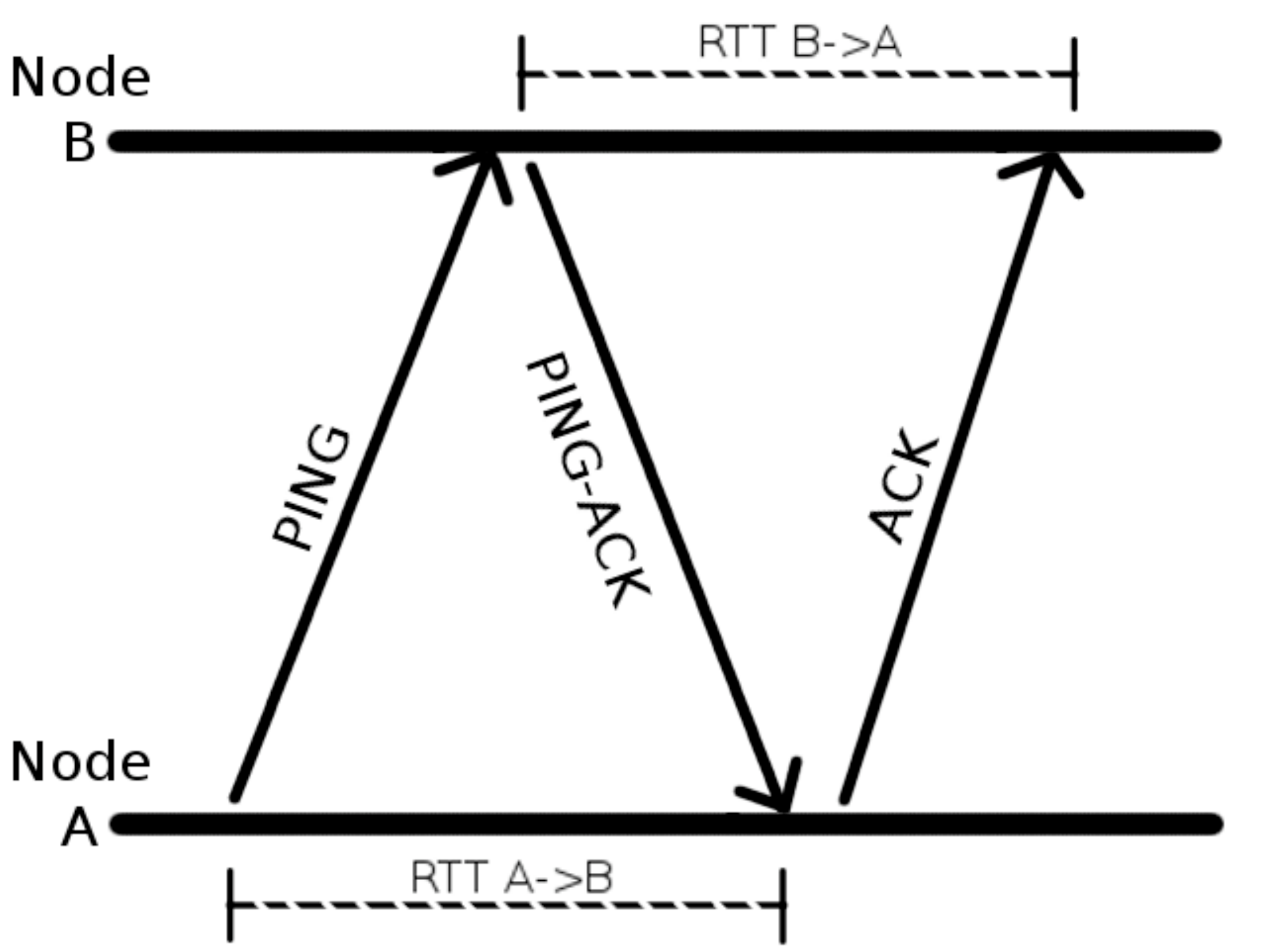} 
 \caption{Sequência de troca de mensagens no TWP}\label{fig:twpseq}
\end{figure}

O \textit{Three-Way Ping} (TWP) funciona da seguinte maneira: o nó A envia
periodicamente solicitações ao nó B, utilizando mensagens UDP. A
solicitação inicial é denominada PING, ao passo que o nó B responderá um
PING-ACK. Ao receber a resposta do nó B, o nó A devolve um ACK para
completar a rodada de troca de mensagens deste protocolo.
Assim, o TWP possibilita
que, em um nó A qualquer, calculemos o RTT de A para B como sendo o
intervalo de tempo entre a emissão um PING e o recebimento da resposta
PING-ACK. Para um momento próximo, é também possível
calcular em B o RTT de B para A como sendo o intervalo entre a emissão de
um PING-ACK e o recebimento do ACK, conforme ilustra a
figura~\ref{fig:twpseq}. 
	
Implementamos o TWP em Java, como um software para coleta dos atrasos de
rede no PlanetLab. Cada mensagem do protocolo consiste em um datagrama
UDP com a especificação do tipo de mensagem (PING, PING-ACK ou ACK) e um
número de sequência, para ordenação e identificação única de cada
sequência de três mensagens. Para os identificadores, que cobriram todo o
espaço de endereçamento a eles dedicado algumas vezes durante o
experimento, foi utilizado um mecanismo idêntico ao utilizado pelo
protocolo TCP no processamento de seus números de sequência.
No ato de envio e recebimento das mensagens, cada nó armazena em seu
\textit{log}, em binário, a quíntupla:
(\textit{timestamp},
número de sequência,
tipo de mensagem,
ID do nó origem,
ID do nó destino).

Dessa forma, cada evento de envio ou recebimento ocupa apenas 15 bytes
no \textit{log}. Os IDs dos nós origem e destino não consistem em seus
endereços de rede, mas em um identificador provido pela sua respectiva ordem
em uma tabela \textit{hash} contendo todos os nós e seus respectivos portos.
Assim, após a tabela \textit{hash} ser preenchida com todos os nós do
experimento, obtém-se uma listagem do conjunto de chaves, onde o ID de cada
nó será a posição dele naquele conjunto. Com isso foi possível economizar
espaço no
armazenamento do \textit{log} e, ainda, a utilização dos IDs incrementais
facilitou o processamento posterior.

Para reduzir a quantidade de mensagens trocadas e evitar redundância,
cada nó inicia TWPs para metade do conjunto restante de nós e
responde às solicitações da outra metade do conjunto. 
Para tal utilizamos um iterador circular sobre o conjunto de chaves da
tabela \textit{hash}. Assim, a cada intervalo de comunicação um nó A
somente enviará ou receberá requisições TWP para/de um outro nó B. Isso é
possível pois a sequência das chaves é a mesma em todos os nós. 

\subsection{O processo de coleta}

A operação de todo o sistema é controlada por um nó coordenador, que
centraliza a identificação de todos os nós no início da coleta e
redistribui as informações de inicialização para todos os nós
participantes, determinando como cada nó deverá atuar
durante o experimento.
Os nós participantes armazenam o \textit{log} localmente com um mecanismo
de rotação periódica. Os arquivos que já foram rotacionados são comprimidos e
enviados para o servidor por uma tarefa de fundo (\textit{background}),
controlada por um sistema de escalonamento de transmissões para evitar que
os nós sobrecarreguem o coordenador com acessos simultâneos.
Ao final do experimento, o coordenador envia um comando para os nós
participantes finalizarem a troca de mensagens e enviarem a ele os
resultados restantes.

Numa análise do progresso evolutivo do PlanetLab,
Tang e outros~\cite{tang07-empiricalStudyPlab}
observaram que na maior parte do tempo, mais de 80\% dos nós estavam
\emph{online}, sendo que certos conjuntos de nós apresentavam uma
confiabilidade
maior que os demais. Tipicamente o
\textit{uptime} dos nós do PlanetLab é em média 5 dias~\cite{plab-mttf}. 
A coleta foi realizada por 10 dias em 2009, do dia 27 de maio ao dia 05
de junho. Nesse período, 14 nós foram reiniciados pelo menos uma vez.
Quando um nó era reiniciado ele comunicava-se
com o coordenador para obter a lista dos demais nós
participantes e voltava a trocar requisições TWP com eles.
Entretanto, durante a reinicialização desses 14 nós houve o problema de nova
sincronização de relógios pelo NTP e de identificadores pelos programas de
monitoração, que comprometeram a interpretação dos \emph{logs}.  
Outros 5 nós apresentaram problemas de
contato ao servidor NTP associado.

A premissa da sincronização de relógios não seria importante para uma
análise local dos tempos em uma máquina, mas é essencial para podermos
correlacionar as medições de diversas máquinas ao longo do tempo para as
análises com base temporal. Por esse motivo, optamos por descartar os 19 nós
que tiveram algum problema de sincronização durante o período.
Assim, os logs utilizados neste trabalho ser referem aos 81 nós restantes.

Nos 81 \textit{logs} analisados, ignoramos as três primeiras e as três
últimas horas 
do experimento para evitar valores discrepantes nas perdas e
atrasos, devidos ao início e encerramento dos diversos programas envolvidos
e à sobrecarga do coordenador com as transferências finais.
Acerca da intrusividade da ferramenta, a monitoração gerou um volume de
tráfego de cerca de 2,2 Kbps por máquina (com 10 segundos entre TWPs). Tal
volume é pouco significativo para a rede do PlanetLab, onde os enlaces de
última milha nunca são inferiores a 1 Mbps. Quanto à carga de
processamento gerada nos nós, essa foi sempre inferior a 1\% (CPU
\textit{utilization}).


\section{Análise dos dados coletados}
\label{sec:resultados}

O tamanho médio dos \textit{logs} colhidos foi de 85.000
TWP/nó,
o que leva a cerca de 255.000 mensagens trocadas entre cada par de nós.
Com 81 nós sendo considerados na análise final temos um total de 3.240
enlaces bidirecionais e, consequentemente, pouco mais de 826 milhões de
mensagens trocadas durante os 10 dias da coleta.
Ao todo, os 81 \textit{logs} binários ocuparam 20~GiB de
espaço em disco (4~GiB, com compressão 7Zip).
O \textit{log} completo, bem como os programas de medição e coleta dos TWP
podem ser obtidos contactando-se o primeiro autor.


Conforme discutido na seção~\ref{sec:medicao} e ilustrado na
fig.~\ref{fig:twpseq}, para dois nós A e B, os tempos de ida-e-volta entre
eles são definidos como o tempo decorrido entre o envio e a recepção de
duas das mensagens do \emph{three-way ping} (PING/PING-ACK ou PING-ACK/ACK,
conforme o caso).

Caso ambos os nós A e B estejam com seus relógios sincronizados, é
possível também calcular os atrasos individuais se estes forem de ordem
de grandeza superior ao escorregamento dos relógios (\textit{clock-drift})%
\footnote{O termo escorregamento é
adotado pela RNP como a tradução mais indicada para \emph{drift} nesse caso
--- vide\\\url{http://ntp.br/NTP/MenuNTPVocabulario}}%
.
Os relógios dos nós do PlanetLab apresentam valores reduzidos de
\emph{drift} em relação a seus servidores NTP:
60\% apresentam um erro de menos de 2~ms e 40\%
apresentam um erro máximo estimado pelo NTP de menos de
10~ms~\cite{pathak2008measurement}. 
Com base nessa premissa, 
o atraso direto de A para B é calculado subtraindo-se o tempo de chegada da
solicitação PING em B do seu tempo de envio em A. O atraso inverso também
poderá ser calculado ao subtrair-se o tempo de chegada da resposta PING-ACK
em A do seu tempo de envio em B. Ademais, uma estimativa da precisão do
atraso direto pode ser realizada se for considerado também o atraso direto
disponibilizado pelos tempos de envio e chegada da confirmação final ACK.
Estes atrasos individuais serão utilizados na análise de assimetria na
seção~\ref{sec:assimetria}.

Consideramos como mensagens perdidas aquelas que aparecem no \emph{log} da
máquina origem mas que não são registradas na máquina de destino 
(identificados pelo número de sequência de cada mensagem). Nesses casos,
o RTT em pelo menos um dos sentidos não pode ser calculado.

\subsection{Evolução diária}

\begin{figure*}[htb!]
 \centering
 \subfigure[RTT médio]{
 \includegraphics[width=0.9\columnwidth]{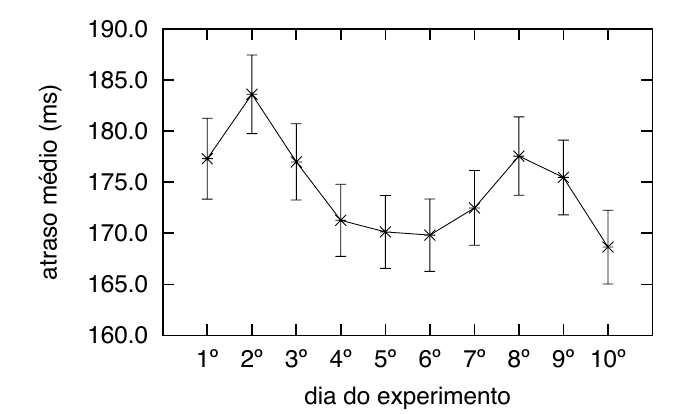}
 \label{fig:log-dailymeanrtt}
 }
 \subfigure[Variação do RTT (CV)]{
 \includegraphics[width=0.9\columnwidth]{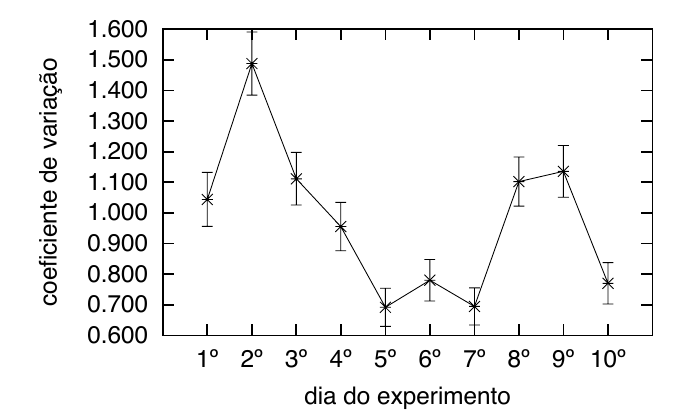}
 \label{fig:log-dailycvrtt}
 }
 \subfigure[RTT mediano]{
 \includegraphics[width=0.9\columnwidth]{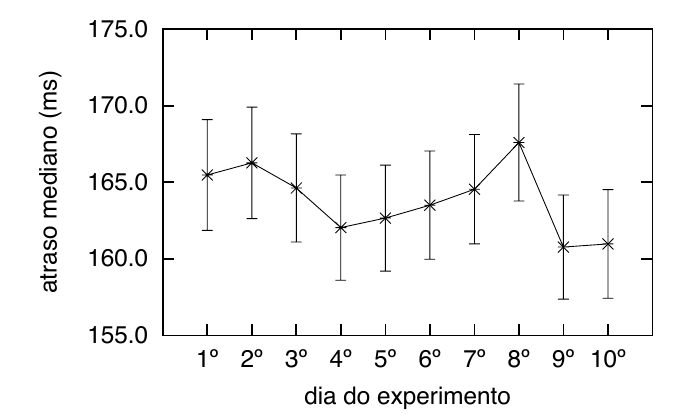}
 \label{fig:log-dailymedian}
 }
 \subfigure[Perdas]{
 \includegraphics[width=0.9\columnwidth]{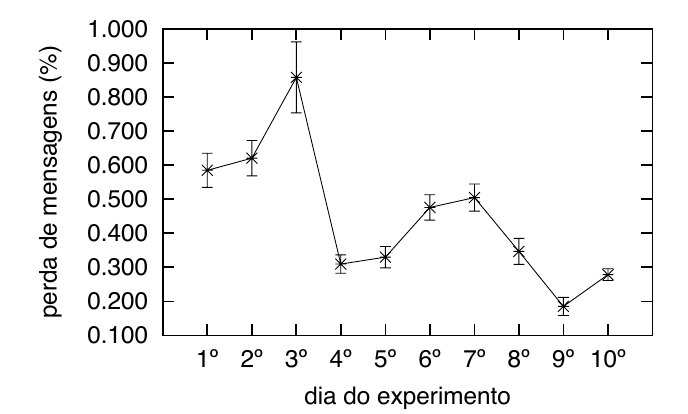}
 \label{fig:log-dailyloss}
 }
 \caption{Evolução diária das medições}\label{fig:log-daily}
\end{figure*}

Como mencionado anteriormente, a coleta foi realizada dos dias
27/05/2009 a 05/06/2009,
incluindo portanto um final de semana (quarto e quinto dias da coleta).
A figura~\ref{fig:log-daily} apresenta as médias diárias para algumas
grandezas obtidas a partir do \textit{log}.
Nos quatro casos, as barras verticais 
indicam os intervalos com 99\% de confiança para a grandeza calculada.

A figura~\ref{fig:log-dailymeanrtt}
exibe a evolução diária do RTT médio ao longo
do \textit{log}. Nela é possível observar que do segundo ao sexto dia houve um
período de queda no RTT médio, que foi de 183~ms a 169~ms, incluindo
um fim de semana. Após este período observa-se uma breve subida do sétimo
ao oitavo
dia para o valor de 177~ms e uma nova queda até o fim da coleta no décimo dia,
alcançando então o menor valor para o RTT médio: 168~ms.
Já a mediana, exibida na figura~\ref{fig:log-dailymedian}, 
manteve-se ligeiramente inferior à média, indicando uma concentração maior em
valores mais baixos.

A variabilidade das medições é ilustrada na
figura~\ref{fig:log-dailycvrtt}, que exibe a evolução diária da
variabilidade média dos atrasos ao longo do \textit{log}.
Nela é possível observar que, tal como observado no RTT médio, do segundo
ao oitavo
dia houve um período de queda na variabilidade, seguido de uma subida do
oitavo
ao nono dia. Ou seja, ao mesmo tempo que o RTT médio decaiu, observou-se uma
maior homogeneidade no coeficiente de variação dos atrasos do referido
período. 

Finalmente, a
figura~\ref{fig:log-dailyloss} exibe a evolução diária da perda média
de mensagens ao longo do \textit{log}. Observamos que a perda de
mensagens apresentou uma maior variabilidade nos 3 primeiros dias da
coleta, passando a demostrar valores mais homogêneos a partir do quarto dia. O
terceiro dia apresentou o maior valor global de 0,86\% para a perda média de
mensagens e o nono dia apresentou o menor valor global de 0,18\%.

É interessante
observar que as médias diárias, calculadas considerando todos os 81 pares,
apresentam quedas durante o final de semana para todas as grandezas
ilustradas na figura~\ref{fig:log-daily}. Se aceitarmos que o tráfego na
Internet pode ser mais reduzido durante os finais de semana, a variação
observada pode ser atribuída à redução das filas nos roteadores naquele
período (e, consequentemente, a redução da fração da latência associada).

\subsection{Distribuição Acumulada dos RTTs}

Uma outra forma de interpretar os dados é determinando como os tempos de
ida-e-volta e as perdas se comportam 
para cada par de
máquinas participantes. Para isso, calculamos as métricas de interesse
separadamente para cada par e apresentamos aqui as distribuições acumuladas
obtidas.

\begin{figure*}[htb!]
 \centering
 \subfigure[RTT médio (ms)]{
 \includegraphics[width=0.9\columnwidth]{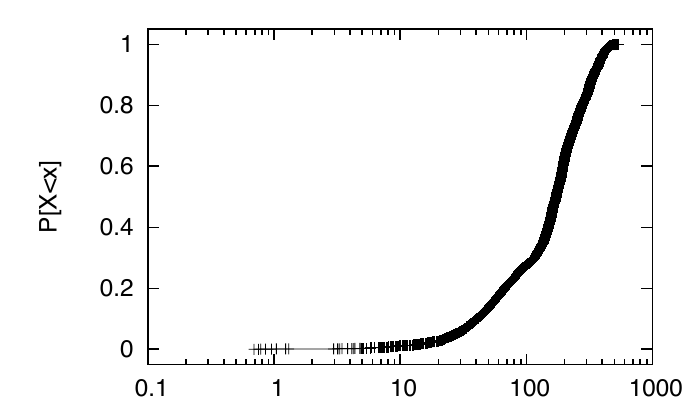}
 \label{fig:log-meanrtt}
 }
  \subfigure[RTT 99-quantil (ms)]{
 \includegraphics[width=0.9\columnwidth]{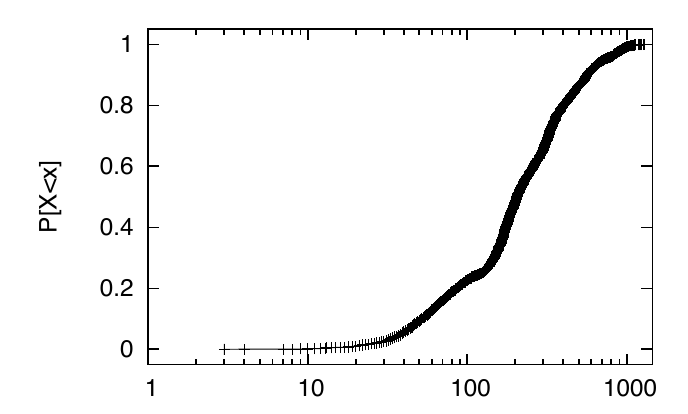}
 \label{fig:log-q99}
 }
 \subfigure[Coeficiente de Variação (CV)]{
 \includegraphics[width=0.9\columnwidth]{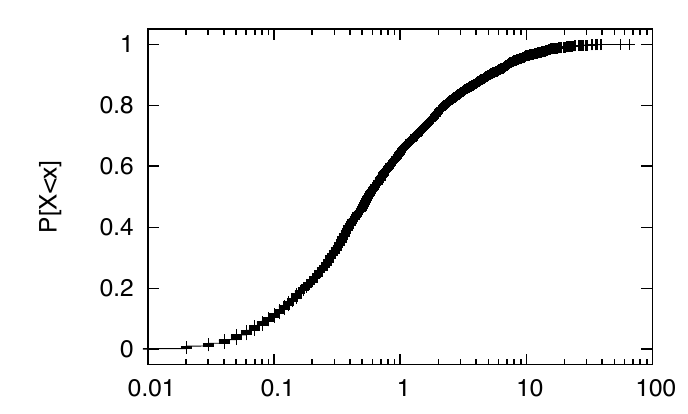}
 \label{fig:log-cvrtt}
 }
 \subfigure[Perdas (\%)]{
 \includegraphics[width=0.9\columnwidth]{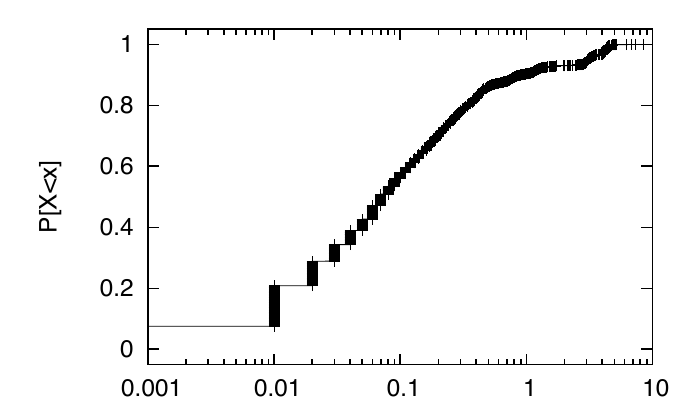}
 \label{fig:log-loss}
 }
 \caption{Distribuições acumuladas das medições}\label{fig:log-cdfs}
\end{figure*}

Para o RTT, foi obtida uma média global de 181~ms.
Conforme pode ser observado na
figura~\ref{fig:log-meanrtt}, apenas 40\% dos pares apresentaram médias
acima de 181~ms, com um máximo de 537~ms para o RTT médio. Se observarmos
os resultados para o 99-quantil na figura~\ref{fig:log-q99}, veremos que em 90\% dos pares de máquinas
considerados, apenas um por cento dos RTTs medidos ficou acima de 500~ms.

O coeficiente de variação (CV) é apresentado na figura~\ref{fig:log-cvrtt} e
indica a variabilidade dos dados. É calculado pelo desvio padrão dividido
pela média, de maneira que quanto menor o CV, mais homogêneo é o conjunto
de dados.
Foram observadas altas variabilidades nos atrasos das mensagens para a
maioria dos pares. Uma fração pequena dos pares (menos de 5\%) 
apresentaram CVs abaixo de 0,05; 50\% ficaram abaixo de 0,56, e 50\%
ficaram entre 0,22 e 1,74.

Na figura~\ref{fig:log-loss}
podemos observar que, em geral, a taxa de perda de mensagens foi baixa. A
taxa de perda média foi de 0,41\% apenas, cerca de 60\% dos pares de máquinas
experimentaram perdas inferiores a 0,1\% e
90\% dos nós apresentaram perdas abaixo de 0,90\%, com um
máximo de 8,6\% de perda de mensagens.

Quanto aos valores extremos (máximos e mínimos) do RTT, cujos gráficos não
foram incluídos por questões de espaço, observamos que
a média dos RTTs mínimos foi de 145~ms, com 75\% dos nós abaixo de 198~ms e
95\% abaixo de 330~ms. O mínimo absoluto foi de 1~ms (entre nós de uma
mesma rede) e o maior valor observado para o RTT mínimo foi de 409~ms.
Já a média dos RTTs máximos foi de 27 segundos, com 75\% dos nós abaixo
de 32 segundos e 95\% abaixo de 112 segundos. O máximo absoluto foi de 170 segundos, ou
seja, 2,8 minutos de atraso. Esses casos de valores muito altos não se
encontravam em rajadas, mas, como mensagens que aparentemente se separaram
do fluxo usual e acabaram chegando ao destino muito mais tarde que
aquelas enviadas imediatamente antes e depois delas.

\subsection{Caracterização da Distribuição dos Atrasos}

Um dos objetivos deste trabalho é identificar a melhor distribuição
estatística que modelasse os atrasos medidos. Tal distribuição poderia ser
utilizada em trabalhos que necessitem gerar valores sintéticos de RTT que
acompanhem uma distribuição realista.
Para determinar a distribuição utilizamos o software
Minitab\footnote{\url{http://www.minitab.com/}}.
Devido a suas limitações em termos da
quantidade máxima de linhas permitidas e do tempo necessário para o
processamento, dos 551 milhões de medidas de RTT coletados obtivemos uma
amostra aleatória de 0,1\%, ou 551 mil medidas.

A distribuição estatística dos dados pode ser encontrada através de
testes padrões para determinar o grau de adequação
(\textit{goodness of fit}): Chi-square,
Kolmogorov-Smirnov (KS) e Anderson-Darling (AD)~\cite{stephens1974edf}, 
dos quais o Minitab utiliza esse último.

Há uma grande quantidade de
distribuições estatísticas que consistiriam em candidatos a modelo de
distribuição dos
dados, das quais a tabela~\ref{tab:goodnessoffit} lista as distribuições que se 
mostraram mais apropriadas. 
Na segunda coluna, o coeficiente AD fornece-nos uma medida 
da qualidade do ajuste. Ao compararmos o ajuste das 
distribuições para o conjunto de dados, a distribuição com o menor
valor de Anderson-Darling oferecerá o melhor ajuste (\textit{best fit}). 
Porém é necessário que ela rejeite a hipótese nula de que os dados têm distribuição normal, de maneira que o \textit{p-value} deve ser menor ou igual a 0,05.

\begin{table}[htb!]
	\begin{center}
		\begin{tabular}{|l|c|c|}
\hline
\em \textbf{Distribuição} & \em \textbf{AD} & \em \textbf{P-Value} \\
\hline
3-Par Lognormal	        &	203	& *	\\
Gamma			&	208	& $<$ 0,005	\\
3-Par Loglogistic	&	214	& *	\\
Lognormal		&	267	& $<$ 0,005	\\
3-Par Weibull	        &	273	& $<$ 0,005	\\
Weibull			&	472	& $<$ 0,010	\\
Normal			&	986	& $<$ 0,005	\\
2-Par Exponential	& 	2376	& $<$ 0,010	\\
\hline
		\end{tabular} 	
	\end{center}
        \caption{Qualidade do ajuste da distribuição dos atrasos}
        \label{tab:goodnessoffit}	
\end{table}

\begin{figure*}[htb!]
 \centering
 \includegraphics[width=0.9\textwidth]{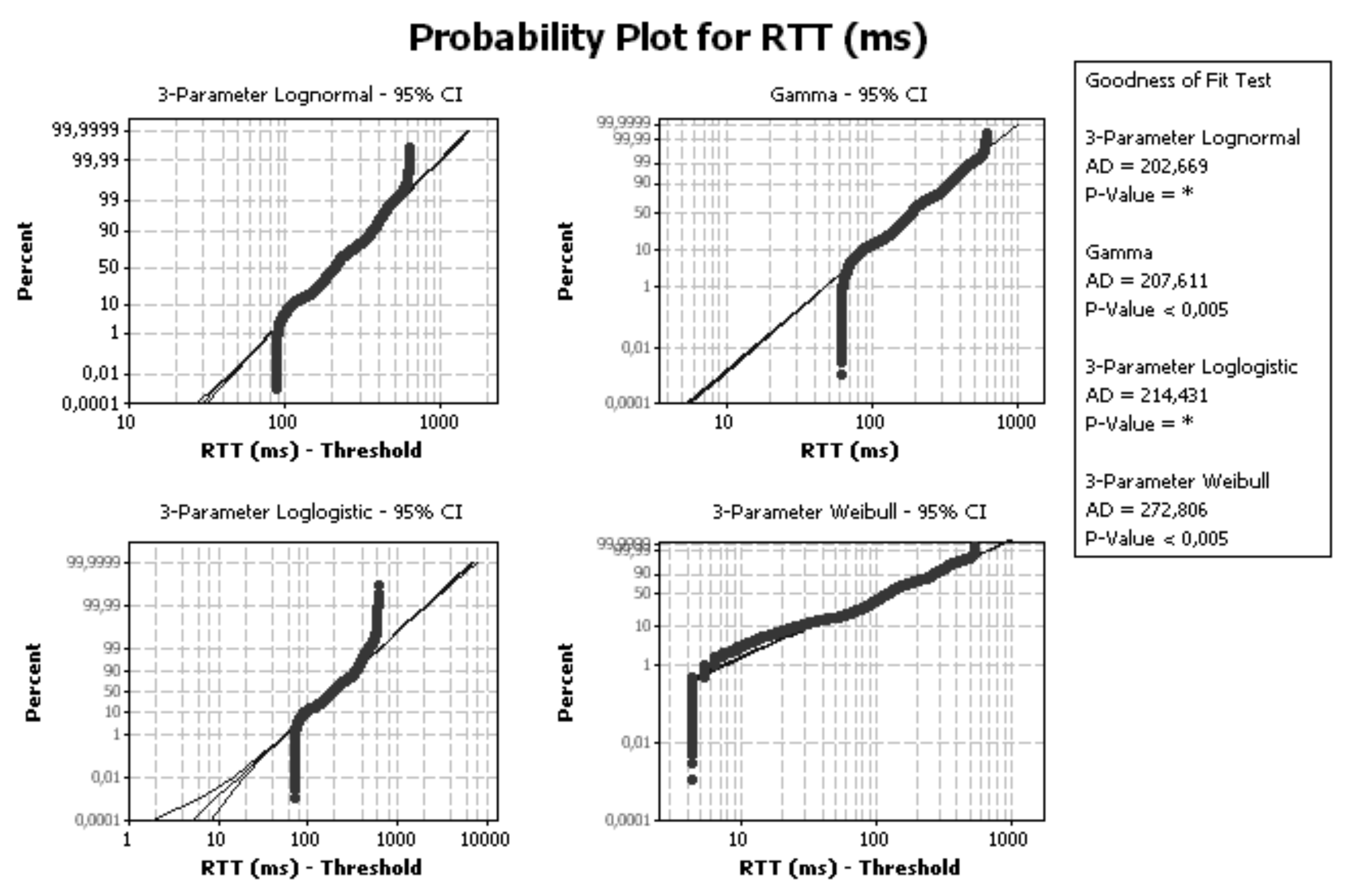}
 \caption{Gráfico de probabilidades das distribuições}\label{fig:bestfit}
\end{figure*}

A figura~\ref{fig:bestfit}
exibe os gráficos de probabilidade de ajuste daquelas
distribuições, consistindo numa técnica para avaliar se o conjunto de dados
segue ou não uma dada distribuição. Os dados são plotados sobre a
distribuição candidata de maneira que os pontos devam formar
aproximadamente uma linha reta e desvios na reta indicam desvios da
distribuição especificada~\cite{croarkin2002engineering}. Sendo assim,
nosso critério para a distribuição com melhor ajuste é aquela com o gráfico
de probabilidade mais linear. O coeficiente de correlação AD provê uma
medida da linearidade do gráfico, assim como uma medida do quão bem a
distribuição ajusta-se aos dados. 

Ao observar a tabela~\ref{tab:goodnessoffit}  e a figura~\ref{fig:bestfit}
verificamos que as distribuições 
Lognormal de 3 parâmetros e
Gamma foram as que melhor se ajustaram aos atrasos observados. 
Porém, com a Lognormal de 3 parâmetros não foi possível rejeitar a hipótese de normalidade. 
Além disso, das duas, a distribuição Gamma foi a que melhor se ajustou aos
extremos (atrasos $<$100~ms e $>$500~ms).
Portanto, propomos a utilização da
distribuição Gamma como um modelo de
distribuição apropriado para a modelagem de atrasos na rede, 
provendo um balanço entre melhor ajuste
 e simplicidade.

A distribuição Gamma é uma distribuição de probabilidade contínua que
possui dois parâmetros: um parâmetro de escala $\Theta$ e um parâmetro de
forma $\lambda$. Ela é frequentemente utilizada para modelar tempos de
espera e aparece naturalmente em processos para os quais os tempos de
espera entre eventos com distribuição \textit{Poisson} são relevantes. Ela
pertence à família das distribuições exponenciais com parâmetros naturais
$\lambda$-1 e -1/$\Theta$. A sua função característica $\varphi$x(t) é
$(1-\Theta it)^{-\lambda}$, em que t é o argumento da função característica
e i é a raiz quadrada de menos um. 

\begin{table*}[htb!]
	\caption{Estimativas dos parâmetros da distribuição dos atrasos\label{tab:paramfit}}	
	\begin{center}
		\begin{tabular}{|l|c|c|c|c|}
\hline
\em \textbf{Distribuição} & \em \textbf{Localização $\mu$} & \em \textbf{Forma $\lambda$} & \em \textbf{Escala $\Theta$} & \em \textbf{Limiar} \\
\hline
Gamma				& - 			& 4,63062		& 43,16537		& - \\
3-Par Lognormal		& 5,34111		& -			& 0,41280		& -27,12915 \\
3-Parameter Loglogistic		& 5,25100		& -			& 0,26255		& -9,23031 \\
3-Parameter Weibull		& -			& 1,51208		& 158,33886		& 56,63291 \\
\hline
		\end{tabular} 	
	\end{center}
\end{table*}

A tabela~\ref{tab:paramfit} exibe as estimativas dos parâmetros das
distribuições para modelagem do atraso de mensagens.
Através dela é possível configurar as
distribuições para exibirem atrasos de rede compatíveis com o PlanetLab.
Segundo Silva e outros, a geração de cargas sintéticas mais realistas
também possibilitaria analisar com mais precisão aplicações em contextos
semelhantes ao do \textit{log} coletado, de maneira a facilitar o projeto, implementação e validação de novos protocolos de rede \cite{silva2009livestreaming}.
Desta maneira, este trabalho fornece tanto uma carga real quanto as
distribuições estatísticas mais apropriadas para a modelagem dos atrasos de
rede através de uma carga sintética. 

\begin{figure*}[htb!]
 \centering
 \includegraphics[width=0.8\textwidth]{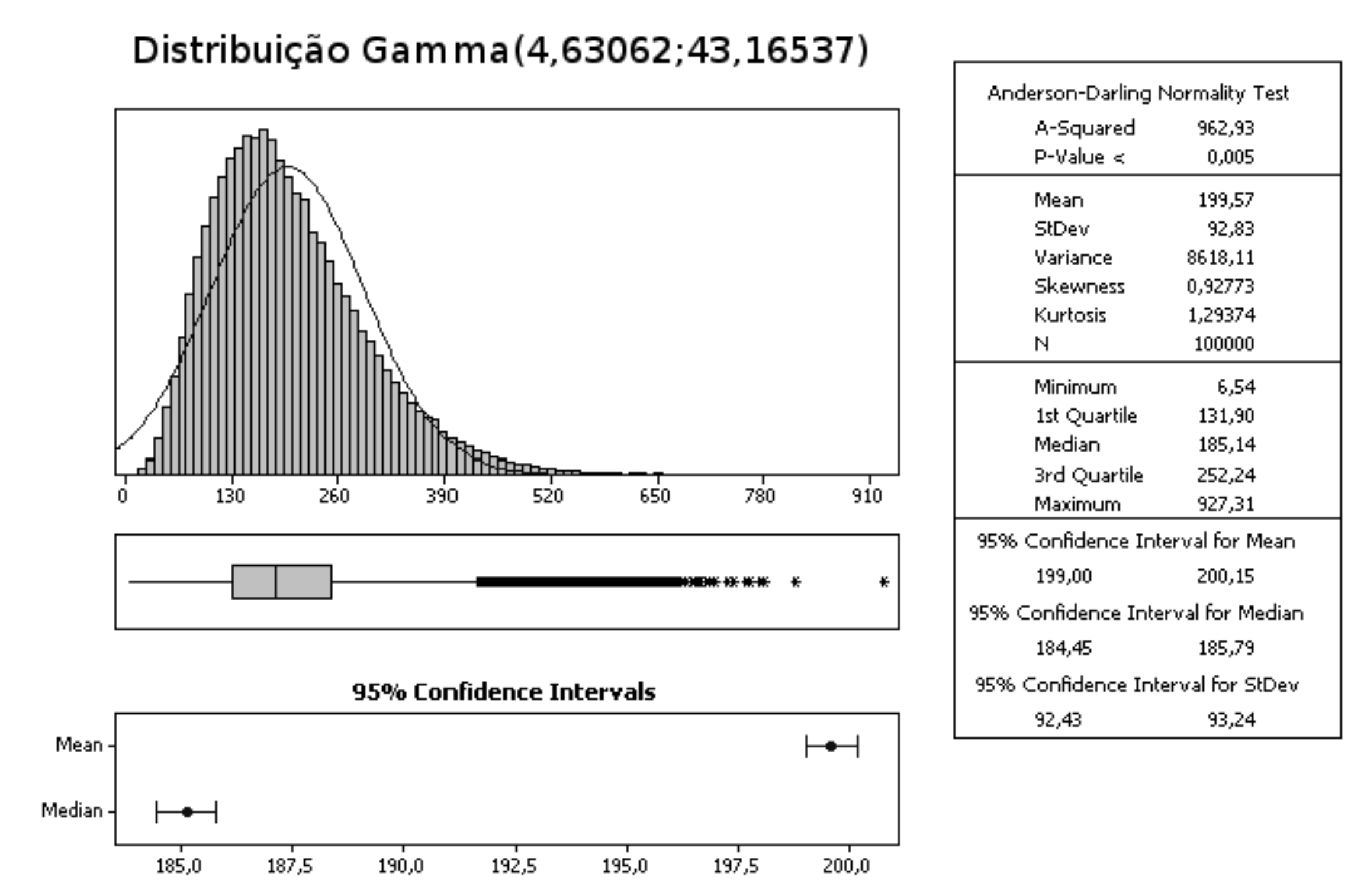}
 \caption{Distribuição Gamma para o RTT do PlanetLab}\label{fig:dist-gamma}
\end{figure*}

A figura~\ref{fig:dist-gamma} exibe as estatísticas descritivas da
distribuição Gamma configurada com os parâmetros de melhor ajuste ao RTT do
PlanetLab. Sua obliquidade positiva (\textit{skewness}=0,927) faz com que
sua massa seja concentrada à esquerda e exiba uma cauda mais longa à
direita, demonstrando uma assimetria na dispersão dos dados em relação à
curva normal (curva sobre o histograma). A curtose também positiva
(\textit{kurtosis}=1,294) caracteriza um menor achatamento da curva em
relação à normal, ou seja, exibe um afunilamento com um pico mais alto e
uma maior concentração que a normal. Consequentemente, a distribuição Gamma
assim configurada possui caudas mais prolongadas, onde é relativamente
fácil obter valores que se afastam da média a vários múltiplos do desvio
padrão. Observe ainda que a média exibida é de 199~ms, ou seja,
ligeiramente diferente da média global exibida pelo PlanetLab de 181~ms
(isto é devido a imperfeições no ajuste da modelagem). 

\subsection{Assimetria dos Atrasos}\label{sec:assimetria}

Em uma análise do progresso evolutivo do RTT (\textit{round-trip time})
médio do PlanetLab, Tang e outros observaram uma
considerável variação no RTT entre pares de nós (de 150 a 550~ms), mesmo
entre curtos períodos de tempo (15 minutos)~\cite{tang07-empiricalStudyPlab}. Isso apresenta um desafio
significativo aos modelos topológicos que pressupõem uma latência
fim-a-fim estática entre pares de nós. Em recente análise da assimetria
dos atrasos na Internet~\cite{pathak2008measurement}, Pathak e outros realizaram um
estudo abrangente no qual mediram a severidade da assimetria no
atraso entre diversos pares de nós do PlanetLab. Seus resultados estão de
acordo com os de Tang e outros, sendo observado que a
assimetria dos atrasos prevalece na grande maioria dos nós, podendo ser
em parte atribuída à assimetria nas rotas e em parte a congestionamentos
transientes. Observaram também que a assimetria é dinâmica, ou seja, à
medida que o tempo progride, a assimetria dos atrasos varia. Além
disso, notaram que redes comerciais exibem níveis mais altos de
assimetria que redes educacionais e de pesquisa.

Visto que a assimetria dos atrasos na Internet é uma realidade, conduzimos
uma análise da assimetria do \textit{log} coletado. Utilizamos a métrica
\textit{assimetria relativa}, que consiste no módulo da diferença entre os
atrasos direto e inverso normalizado pelo atraso mínimo, conforme
representado na equação~\ref{form:assimetria-relativa}. Naquela equação,
quanto mais próximo de zero for o resultado, mais simétrico será e
consequentemente, quanto maior for o valor encontrado, maior será a
assimetria. 

\begin{equation}\label{form:assimetria-relativa}
 assimetria = \frac{| atraso_{AB} - atraso_{BA} |}{min(atraso_{AB},atraso_{BA})}
\end{equation}

Por exemplo, considere que em determinado momento o RTT observado num par
de nós foi de 150~ms e o atraso direto de 60~ms. Portanto, o atraso inverso
foi de 90~ms o que representa uma assimetria de 30~ms entre esses dois
atrasos. Com 60~ms de atraso direto e 90~ms de atraso inverso, temos uma
assimetria relativa de abs(60-90)/60=0,50. Isso significa que o sentido
com maior atraso apresentou um acréscimo de 50\% em relação ao valor do
atraso no outro sentido.

\begin{table*}[!htb]
\caption{Estatísticas da assimetria relativa dos atrasos}\label{table:assimetria}
  \small
\begin{center}
\begin{tabular}{|l|c|c|c|c|c|c|c|c|c|}
\hline
\em \textbf{Média} & \em \textbf{Desvio Padrão} & \em \textbf{CV} & \em \textbf{$Q_{25}$} & \em \textbf{Mediana} & \em \textbf{$Q_{75}$} & \em \textbf{$Q_{90}$} & \em \textbf{$Q_{95}$} & \em \textbf{$Q_{99}$} \\
\hline 
0,947 & 8,78 & 9,27 & 0,08 & 0,20 & 0,56 & 1,62 & 3,08 & 12,07 \\
\hline
\end{tabular}
\end{center}
\end{table*}

\begin{figure}[htb!]
 \centering
 \includegraphics[width=0.9\columnwidth]{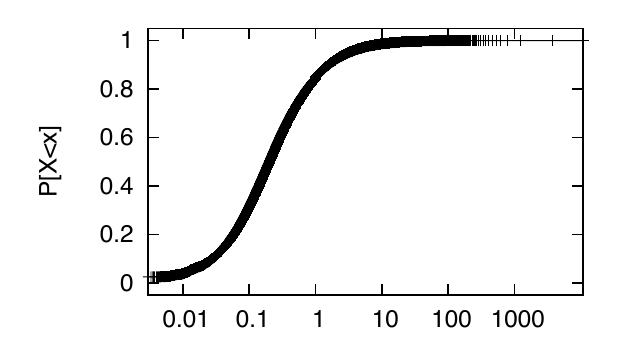}
 \caption{Distribuição acumulada da assimetria relativa dos atrasos}\label{fig:assimetria-resumo}
\end{figure}

Os resultados obtidos para esta métrica estão expressos na tabela~\ref{table:assimetria} e na figura~\ref{fig:assimetria-resumo}, onde podemos observar que a assimetria média do PlanetLab é alta (0,947), porém com uma alta variabilidade (9,27). Mesmo a média calculada com a exclusão de 5\% dos valores discrepantes ainda pode ser considerada alta, com um valor de 0,409.
Metade dos nós apresentou uma assimetria relativa entre 0,08 e 0,56 e 95\% deles apresentaram uma assimetria abaixo de 3,08.

\subsection{Clusterização dos Enlaces}

Realizamos também uma clusterização dos enlaces de acordo com suas
estatísticas descritivas (atraso médio, seus quantis, variabilidade do atraso e \% de perda), buscando separá-los em 
grupos que, para estas métricas, possuam um comportamento aproximado.
Os canais entre duas máquinas A e B foram considerados separadamente em
cada sentido, isto é, as medidas de RTT de A para B foram consideradas como
um canal de medição e as medidas de B para A como outro canal independente.

\begin{table}[htb!]
\caption{Estatísticas da clusterização\label{tab:clustering-statistics}}
	\begin{center}
		\begin{tabular}{|c|c|c|c|c|}
\hline
\em \textbf{Cluster} & \em \textbf{Enlaces} & \multicolumn{2}{c|}{\em \textbf{RTT}}  & \em \textbf{Taxa de} \\
\em \textbf{Cluster} & \em \textbf{(\%)}    & \em \textbf{Média} &  \em \textbf{CV} & \em \textbf{Perda (\%)}  \\
\hline                                                                                       
c1 	& 21\% &  49 ms	& 1,12 & 0,22\%	\\
c2 	& 21\% & 131 ms & 6,37 & 0,40\%	\\
c3 	& 24\% & 167 ms	& 0,33 & 0,30\%	\\
c4 	& 21\% & 269 ms & 0,96 & 0,12\%	\\
c5 	& 13\% & 358 ms & 0,44 & 1,40\%	\\
\hline
Global 	& 100\%& 181 ms & 1,89 & 0,41\% \\
\hline
		\end{tabular} 	
	\end{center}
\end{table} 

Para a clusterização utilizamos o software
\textit{Weka}\footnote{\url{http://www.cs.waikato.ac.nz/ml/weka/}}
 com o algoritmo de clusterização
\textit{Expectation-Maximization} (EM), donde chegamos às 5 instâncias de 
\textit{clusters} listadas na tabela~\ref{tab:clustering-statistics}.
Observamos que o cluster 1 apresentou um baixo RTT médio e baixa perda de
mensagens, ao contrário do cluster 5, que apresentou atrasos relativamente
bem maiores e uma maior taxa de perda de mensagens. O cluster 2 mostrou uma
grande variabilidade nos dados, dado seu alto CV, ao passo que o cluster 3
teve a menor variabilidade. Acerca do cluster 4, podemos dizer que ele
apresentou um comportamento próximo da média global do PlanetLab.

\begin{figure*}[htb!]
 \centering
 \subfigure[cluster 1]{
 \includegraphics[width=0.9\columnwidth]{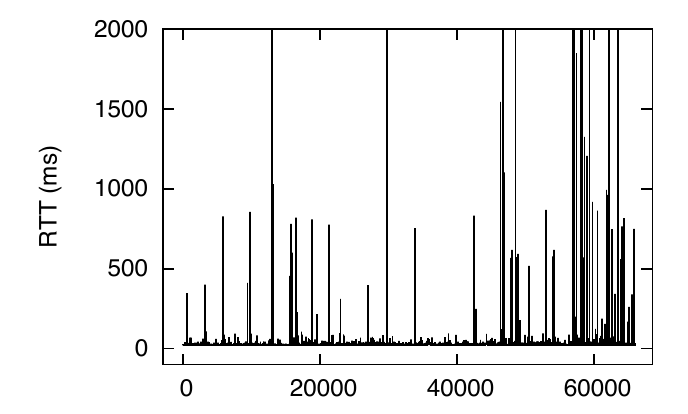}
 \label{fig:seriesc1}
 }
 \subfigure[cluster 2]{
 \includegraphics[width=0.9\columnwidth]{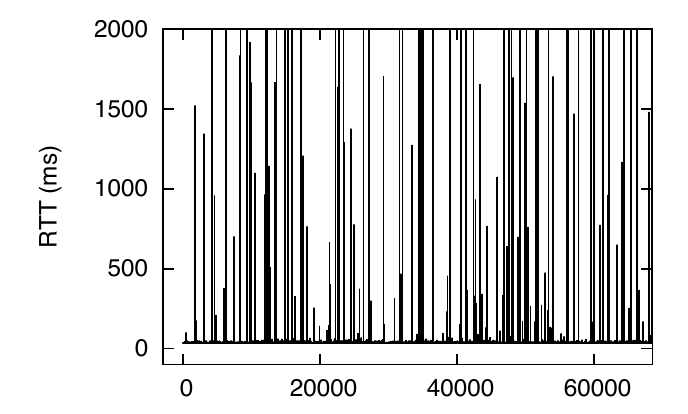}
 \label{fig:seriesc2}
 }
 \subfigure[cluster 3]{
 \includegraphics[width=0.9\columnwidth]{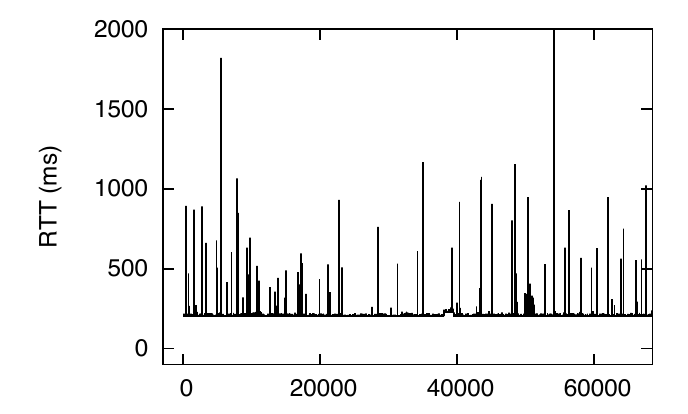}
 \label{fig:seriesc3}
 }
 \subfigure[cluster 4]{
 \includegraphics[width=0.9\columnwidth]{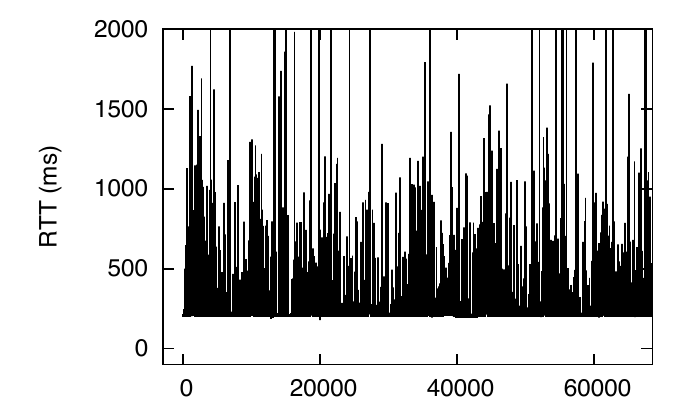}
 \label{fig:seriesc4}
 }
 \caption{RTTs ao longo do tempo para os diferentes \textit{clusters}}
 \label{fig:series}
\end{figure*}

A fim de exibir o comportamento típico dos RTTs medidos por um nó, 
selecionamos aleatoriamente uma conexão de cada
cluster e geramos a série temporal dos RTTs medidos para aquela conexão. A
figura~\ref{fig:series} traz os resultados deste procedimento.

Pela figura~\ref{fig:series}, nos exemplos pode-se observar algumas das
características que identificam cada cluster:
em (a) os valores baixos da maioria das medidas de RTT para o par do cluster 1, porém com muitos
picos elevados (alta variabilidade);
em (b) a variabilidade ainda mais elevada do par do cluster 2, ainda com um limite
inferior baixo;
em (c) o menor número de
variações nas medidas do cluster 3, porém já com um limite inferior mais alto e, por fim, 
em (d), para o cluster 4, um valor médio ainda mais alto com significativas variações. 
Um exemplo de série temporal para o cluster 5
não foi incluído por limitações de espaço, mas também apresenta padrão visualmente
diferente dos demais.

Para ilustrar mais claramente as diferenças entre os grupos exibimos abaixo
as distribuições acumuladas, já apresentadas para o \textit{log} completo
mas agora com uma curva para cada um dos \textit{clusters} encontrados.

\begin{figure*}[htb!]
 \centering
 \subfigure[RTT médio (ms)]{
 \includegraphics[width=0.9\columnwidth]{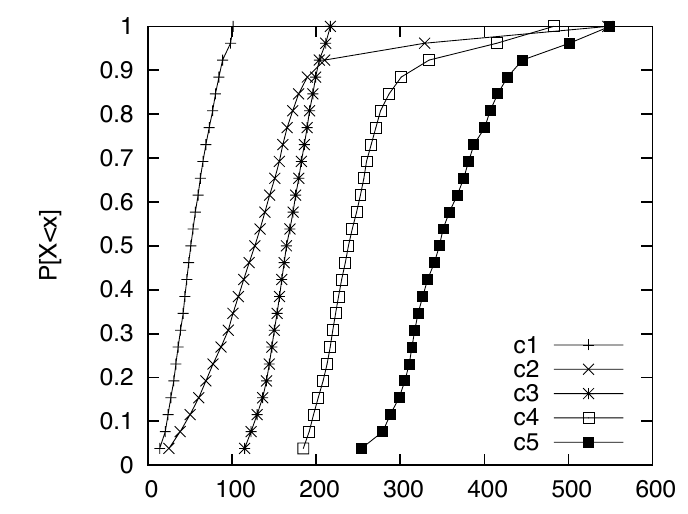}
 \label{fig:clusters-rtt}
 }
 \subfigure[Variação (CV)]{
 \includegraphics[width=0.9\columnwidth]{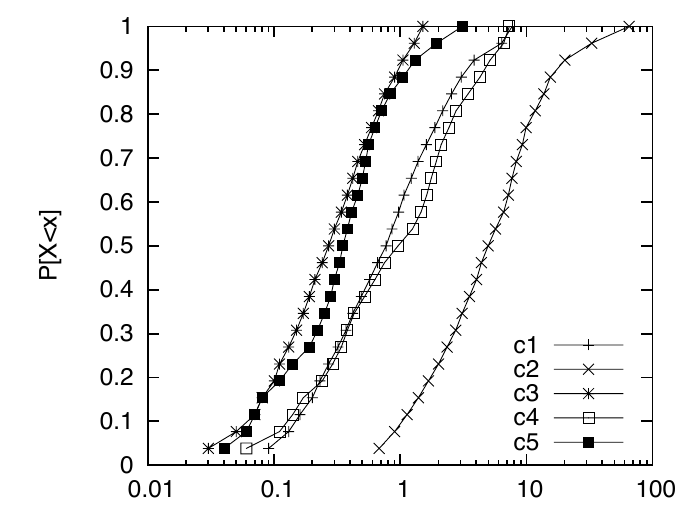}
 \label{fig:clusters-cv}
 }
 \subfigure[Assimetria]{
 \includegraphics[width=0.9\columnwidth]{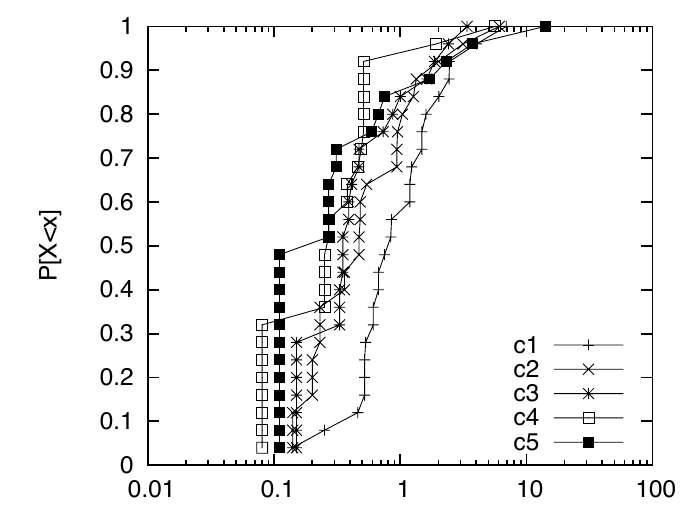}
 \label{fig:clusters-assimetria}
 }
 \subfigure[Perdas (\%)]{
 \includegraphics[width=0.9\columnwidth]{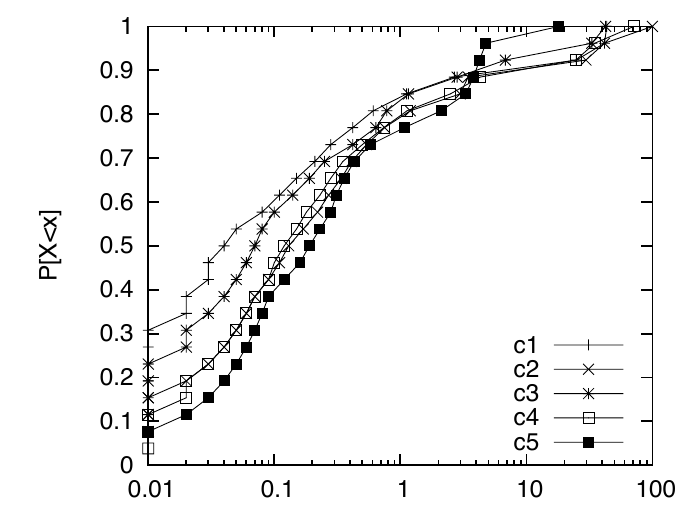}
 \label{fig:clusters-perdas}
 }
 \caption{Análise comparativa dos \textit{clusters}}\label{fig:clusters}
\end{figure*}

Na figura~\ref{fig:clusters}, notamos que as curvas de RTT médio (fig.~\ref{fig:clusters-rtt}) são as que
melhor separam os \textit{clusters}. Nas demais, alguns \textit{clusters} apresentam
distribuições semelhantes, mas ainda assim é possível observar as
separações. As distribuições de perdas, apesar de parecerem próximas, se
diferenciam mais claramente nas duas extremidades, pela quantidade de nós
que experimentam poucas perdas em cada cluster e pelos valores
máximos de perdas em cada caso.

Quanto à topologia, analisamos a composição da distribuição geográfica dos clusters. Observamos que o cluster 1 é populado principalmente por nós de um mesmo continente, seja ele Europa ou América do Norte. enquanto que o cluster 2 é caracterizado por nós do Oriente Médio ou Mediterrâneo. Já o cluster 3 é constituído na maior parte por pares de nós dos continentes Europa com América do Norte. O cluster 4 é composto essencialmente por pares de nós da América do Sul com outro continente e o cluster 5 por pares de nós da Ásia com outro continente.

	Finalmente, como os enlaces em cada cluster possuem características próximas, seria de
se esperar que as medidas dos canais de A para B e de B para A fossem
classificados no mesmo cluster. 
Como as medidas são obtidas pelo
\textit{Three-Way Ping}, normalmente, um par de medidas (uma em cada sentido)
é obtido de cada vez. 

	Usualmente, as duas medições de RTT em um TWP tendem a ser  
semelhantes, dado que cada uma envolve uma mensagem em cada sentido.  
Entretanto, oscilações rápidas da rede podem ocorrer durante a  
primeira ou a terceira mensagens, levando a valores de RTTs  
ligeiramente diferentes, ou a terceira mensagem pode se perder, o que  
afetaria apenas a segunda medição de RTT. Tais variações, quando  
frequentes, podem levar a casos onde as medidas de A para B sejam  
classificadas em um cluster diferente das de B para A.

\begin{table}[htb!]
\caption{
Distribuição dos canais entre cada par de máquinas A,B.\ \ \ \ \ \ \
A última linha indica a porcentagem dos casos da diagonal principal em
relação ao total em cada \emph{cluster}.
} \label{tab:clustermembership}
\begin{center}
\begin{tabular}{|c|r|r|r|r|r|}
\hline
\emph{cluster} & \multicolumn{5}{c|}{\emph{cluster} de B$\rightarrow$A}\\
\cline{2-6}
de A$\rightarrow$B  & c1 & c2 & c3 & c4 & c5 \\
\hline
 C1      & 1876  & 290   & 14    & 3     & 0     \\ 
 C2      &       & 1078  & 205   & 24    & 7     \\ 
 C3      &       &       & 2084  & 112   & 5     \\ 
 C4      &       &       &       & 1464  & 140   \\ 
 C5      &       &       &       &       & 1332  \\
\hline
\hline
& \multicolumn{5}{c|}{A$\rightarrow$B $\in$ cj $\wedge$ B$\rightarrow$A $\in$ cj}\\ 
\cline{2-6}
& 86\% & 67\%  & 86\% & 84\% & 90\%     \\ 
 \hline
\end{tabular} 	
\end{center}
\end{table}

A tabela~\ref{tab:clustermembership} mostra, para todos os pares A,B,
quantos
foram classificados no mesmo \textit{cluster} (diagonal) e quantos
pares foram divididos entre dois clusters diferentes.
Por exemplo, existem
205 pares em que o canal em um sentido foi classificado em C2 enquanto o
canal no sentido contrário foi classificado em C3. Apenas C1 e C5 não têm
nenhum par compartilhado dessa forma, o que se deve certamente ao fato de
serem os agrupamentos com maiores diferenças entre si.
A última linha resume, para cada \textit{cluster}, a porcentagem dos canais
assinalados àquele \textit{cluster} que formam pares (ambos os sentidos,
A$\rightarrow$B e B$\rightarrow$A,
no mesmo \emph{cluster}).

O cluster 2, 
que teve a maior quantidade de pares com canais também atribuídos a outros grupos, 
foi aquele que teve a maior variabilidade (CV), muito acima
dos demais. Essa maior variabilidade pode ter contribuído para maior
dispersão dos seus elementos.
%
%

\section{Conclusão e trabalhos futuros}

Tempos de ida-e-volta são importantes em diversas situações na Internet:
eles podem, por exemplo, determinar a seleção de certos
pares ou servidores em aplicações distribuídas, afetando a forma como
participantes do sistema se organizam. Por esse motivo, informações sobre 
RTTs são importantes para orientar o projeto de protocolos e como entrada
para sistemas de simulação, entre outros fatores.
Neste trabalho apresentamos um metodologia para medição e caracterização
de tempos de ida-e-volta na Internet.
As medições foram realizadas por um período de dez dias,
com um intervalo entre medições baixo (10 segundos),
utilizando 81 máquinas distribuídas pela Internet em quatro continentes.

Os resultados foram avaliadas em termos dos RTTs médios, sua variabilidade ao
longo do tempo entre um mesmo par de máquinas e entre
pares diferentes, bem como as perdas observadas em cada canal de medição.
O padrão de variação dos RTTs ao longo do tempo foi modelado como uma
distribuição Gamma, cujos par\^{a}metros foram identificados, podendo ser
usada na geração de atrasos em ambientes de simulação, por exemplo. Com base
no uso de relógios sincronizados avaliamos também o atraso unidirecional
entre as diversas máquinas, encontrando uma prevalência significativa de
assimetria nos canais.
Finalmente, uma análise dos \emph{logs} utilizando uma técnica de
clusterização mostrou a existência de cinco grupos, 
caracterizados por diferentes combinações de valor médio de RTT,
taxas de perda e variabilidade das medições, entre outros fatores. A 
inspeção de algumas amostras de cada grupo e a análise das distribuições
acumuladas das diversas métricas em cada caso confirmam as diferenças entre
os grupos e composições inter- e intra-continentes.

Identificamos até o presente momento pelo menos duas linhas para trabalhos  
futuros. Por um lado, pretendemos investigar os fatores de rede que  
levaram à formação dos grupos identificados. Esperamos que isso leve a  
um modelo mais detalhado do comportamento da rede quanto aos tempos de  
ida-e-volta, com caracterização dos \textit{clusters} separadamente.  
Paralelamente, 
pretendemos reproduzir o experimento em maior escala, com mais pontos  
de medição e por períodos mais longos. 
A análise de intervalos maiores ou variáveis se encontram entre os
trabalhos que estamos desenvolvendo no momento, junto com o projeto de uma
versão adaptativa da malha de monitoração que permita a realização de
medidas entre um número maior de máquinas. Para tal, utilizaremos uma
topologia do tipo par-a-par para minimizar a complexidade quadrática da
quantidade de nós intercomunicantes.

\section*{Disponibilidade dos dados}

O \textit{trace} completo, bem como os programas de medição e coleta dos TWP
podem ser obtidos através de contato com o primeiro autor.
Mais informações podem ser obtidas na página
do projeto no qual esta pesquisa está inserida:
{\small
\url{http://www.dcc.ufmg.br/~evaladao/dfd}
}


\label{sec:conclusao}

\section*{Agradecimentos}

Esta pesquisa foi parcialmente financiada pelo Instituto Nacional de Ciência
e Tecnologia para a Web - InWeb (MCT/CNPq 573871/2008-6), pelo CNpq,
pela CAPES e pela FAPEMIG.

\bibliographystyle{plain}
\bibliography{rbresd2010-rtts}

\end{document}